\documentclass{aa}  
\usepackage{color}

\usepackage{graphicx}
\usepackage{txfonts}
\begin{document}

   \title{Radio pulsars resonantly accelerating electrons}

   \author{Zaza N. Osmanov\inst{1,2}
          \and
          Swadesh M. Mahajan \inst{3}
          }

   \institute{School of Physics, Free University of Tbilisi, 0183, Tbilisi, Georgia \
    \email{z.osmanov@freeuni.edu.ge}
              \and
              E. Kharadze Georgian National Astrophysical Observatory, Abastumani, 0301, Georgia
                \and
              Institute for Fusion Studies, The University of Texas at Austin, Austin, TX 78712, USA\
                            \email{mahajan@mail.utexas.edu}
             }

   \date{Received ; accepted }

 
  \abstract
  
   {}
   {Based on the recently demonstrated resonant wave-wave process, it is shown that electrons can be  accelerated to ultra-relativistic energies in the magnetospheres of radio pulsars. The energization occurs via the resonant interaction of the electron wave (described by a Klein-Gordon (KG) equation) moving in unison with an intense electromagnetic (EM) wave; the KG wave/particle continuously draws energy from EM. In a brief recapitulation of the general theory, the high energy (resonantly enhanced) electron states are investigated by solving the KG equation, minimally coupled to the EM field.  }
   {The restricted class of solutions, that propagate in phase with EM radiation (functions only of  $\zeta=\omega t-kz$), are explored to serve as a possible basis for the proposed electron energization in the radio pulsars. }
   {We show that the wave-wave resonant energization mechanism could be operative in a broad class of radio pulsars with periods ranging from milliseconds to the normal values ($\sim 1$ sec); it could drive the magnetospheric electrons to acquire energies from $100$s of TeVs (millisecond pulsars) to $10$ ZeVs (normal pulsars).}
   {}


   \maketitle
%

\section{Introduction} 

In a recent work a new mechanism of wave-particle resonant interaction has been discovered \citep{ma}, that might be a significant process for acceleration of particles. The study of acceleration mechanisms is crucial for understanding the nature and origin of cosmic rays. Earlier research suggested that very high energy (VHE) particles \citep{supern} could be emitted in supernovae events. The Crab Nebula, for instance, was identified as the source of the cosmic rays by \cite{sekido}. The discovery of pulsars in 1967 brought another possible player into the game. In the pioneering work of \cite{gunn}, the role of pulsars in accelerating protons to $10^{18}$eV was established. Since then, a number of authors have investigated pulsars as an energizing source \citep{bednarek,lemoine,or09,or17,screp13,screp15}. It is also strongly believed that extremely high energy cosmic rays may emerge from active galactic nuclei (AGN) \citep{agn1,agn2} and gamma ray bursts \citep{grb1,grb2}. In this context it is worth noting that termination shocks of galactic outflows and star-forming regions are also considered as possible sources for ultra-high energy particles \citep{marc,glbl,owen}.

It is, perhaps, useful to list  the generally invoked  mechanisms that might, potentially, energize the cosmic particles to VHE. The Fermi-type acceleration mechanisms \citep{fermi1,fermi2,catanese} can explain the generation of ultra-high energies. But the Fermi process is very efficient only if the particles have already been pre-accelerated \citep{rieger}.  In contrast, magneto-centrifugal acceleration (MA) does not require pre-acceleration; it can operate on particles starting with lower energies. The high efficiency of the mechanism has been demonstrated in the context of  pulsars \citep{or09,or17,screp13,screp15} and AGN \citep{zev}. The MA, however, works only if magnetic fields are strong enough to ensure that the plasma obeys the "frozen-in condition" forcing the particles follow the co-rotating field lines \citep{or09,or17,screp13,screp15}.

In our recent work \citep{resonAGN} we explored the resonant wave-wave mechanism  (just the one we will be invoking in this paper) of particle energization in the context of the radio-loud  AGN. 

Taking into account the factors limiting the maximum energy (inverse Compton mechanism, synchrotron radiation and curvature emission), we have found that the electrons and protons (in the magnetosphere of a radio-loud AGN) might reach energies of the order of $10^{21}$ eV. It has been shown that the most efficient factor constraining the maximum energies is the radiation field energy density, and consequently inverse Compton scattering becomes significant. In light of the Greisen-Zatsepin-Kuzmin (GZK) limit (a theoretical upper limit of extragalactic cosmic ray protons: $5\times 10^{19}$ eV  \citep{greisen,zk}), a further examination of the efficiency of this new mechanism may be warranted. 

In this paper, however, we turn our attention to pulsars whose magnetospheres are populated with electron-positron pairs. Since many pulsars emit in the radio spectral band, it is interesting to explore whether the resonant energization will, efficiently, operate in this setting.

The paper is organized as follows: in Sec. 2 we summarize the general theory of resonant acceleration\citep{ma,MA-22}. In Sec. 3, we apply the mechanism to pulsars, derive and discuss the principal results. In Sec. 4, we put our results into perspective.

\section{Brief outline of theory}

For a particle with rest mass $m$, momentum, $P = \gamma m\upsilon$, and energy $E= (P^2+m^2)^{1/2}$, the group velocity of the associated KG wave ($\hbar = c = 1$),
\begin{equation}
\label{vg} 
\upsilon_g = \frac{\partial E}{\partial P} = \frac{P}{\left(P^2+m^2\right)^{1/2}},
\end{equation}
tends to unity for $P>>m$. One then, expects, that such a matter wave, could strongly resonate with an EM wave. Moving almost in phase the particle wave could continuously draw energy from the EM wave. 

The analysis of \cite{ma,MA-22} begins with ($\omega = 2\pi f$ is the emission cyclic frequency),
\begin{equation}\label{KG} 
\left(\partial^2_t-\partial_z^2+2qAK_{\perp}\cos\left(\omega t-kz\right)+K_{\perp}^2+m^2+q^2A^2\right)\Psi = 0,
\end{equation}
describing a KG particle wave $(\Psi)$ in the presence of a circularly polarized EM wave
$(A^0 = A^z = 0, \;A^x = A\cos\left(\omega t-kz\right),\; A^y = -A\cos\left(\omega t-kz\right)),$
propagating in the z direction. In Eq. (\ref{KG}), $q$ denotes the charge of the particle and $K_{\perp}$ represents the perpendicular momentum of the KG wave. For the resonant KG-EM system, we seek solutions where $\Psi = \Psi (\omega t-kz)=\Psi (\xi)$. Equation (\ref{KG}), then, converts to an ordinary differential equation
\begin{equation}
\label{mathew} 
\left(\omega^2-k^2\right)\frac{d^2\psi}{d\xi^2}+\left(\mu+\lambda\cos\xi\right)\psi = 0,
\end{equation}
a generic Mathew equation with $\mu = K_{\perp}^2+m^2+q^2A^2$ and $\lambda = 2qAK_{\perp}$. 

Since $\omega$ and $k$ are very close (in vacuum, they are exactly the same), Eq. (\ref{mathew}) is a singular equation. This singular feature ($d^2\Psi/d\xi^2$ must become large to satisfy the equation) is the reason why the KG particle obeying Eq. (\ref{mathew}) must have solutions with high energy $(E)$ and momentum ($K_z$).  The novelty of our mechanism is that the enhancement of electron energy (and the rate of energy gain) depends crucially on the resonance between the two hyperbolic waves, i.e, $E/K_z$. of the electron wave $= \omega/k$ of the EM wave. Both of these are greater than unity (the latter only in a plasma).

It is worth noting that a classical electron is not a hyperbolic wave, but the physics that this theory is based on depends on the resonance of two hyperbolic waves, therefore, the mechanism of particle energization strongly depends on the quantum nature of a particle \citep{MA-22}. {\ bf In fact, we will now show explicitly, that quantum effects are essential for the acceleration process}.

That these high energy solutions are actually accessible, requires further enquiry. This was done by \cite{ma,MA-22} where an expression for the rate of energy gain was derived as
\begin{equation}
\label{rate} 
\frac{dE}{dt}  = \frac{\sqrt {2}\omega^2}{\left(\omega^2-k^2\right)^{1/2}}\frac{\hbar qAK_{\perp}}{\left(m^2+K_{\perp}^2+q^2A^2\right)^{1/2}}\frac{\sin kl}{kl}.
\end{equation}
(where we have restored the $\hbar$ in the numerator) assuming that the particle acceleration takes place in the region $z = \{-l, l\}$. Eq. (\ref{rate}) fully illustrates the resonant character of the acceleration process. Indeed, for small values of $\omega^2-k^2$ the rate of the energy gain becomes very large. One must note :\\
1) the rate of energy gain stems from a quantum effect\\
2) it requires the perpendicular momentum $K_{\perp}$ to be non zero,\\
3) it is resonantly enhanced.\\
To make an estimate for the resonant factor, after restoring dimensions (In the pulsar context, $qA>> mc^2,K_\perp c$), and using the dispersion relation (see  \cite{MA-22}), $\omega^2-k^2c^2 = \omega_p^2 \left(1+q^2A^2/(m^2c^4)\right)^{-1/2}\approx mc^2\omega_p^2/qA$, one obtains
\begin{equation}
\label{rate2} 
 \frac{dE}{dt} \simeq \frac{\sqrt{2}\hbar\omega^2}{\omega_p}\left(\frac{qA}{mc^2}\right)^{1/2}K_{\perp}c. 
\end{equation}
Here, $\omega_p = \sqrt{4\pi nq^2/m}$ is the Langmuir frequency, $n$ denotes the plasma number density. A good estimate for the amplitude of the EM wave is $A = c\sqrt{4\pi cF_r}/\omega$, where $F_r \simeq L_r/(2\pi r^2\theta^2)$ is the radio flux of a pulsar, $\theta\simeq\frac{\Omega r_s}{2c}\left(\frac{R_{lc}}{r_s}\right)^{1/2}$ is the half opening angle of the pulsar's emission beam \citep{mach}, $L_r$ represents the radio luminosity and $r_s\simeq 10$ km is the neutron star's radius.

What we have shown in the paper is a promise- that resonant high energy electron (wave) states exist, and one can qualitatively see an energy-gain pathway through which the electron could be driven to such states.

\section{Theory in the Pulsar Context}

In this section, we apply the aforementioned mechanism to radio pulsars, studying efficiency of the process over a range of physical parameters. 

How long can the accelerating process last?f Until some energy loss mechanisms, that become more efficient  at higher energies, can balance the growth. We will discuss three such loss mechanisms and assess their role in limiting the maximum particle energy.

Synchrotron Radiation: Though the proposed acceleration mechanism was independent of the magnetic field, there does exist a strong magnetic field in the pulsar; the relativistic electrons, moving in such a magnetic field, will experience extreme synchrotron losses, characterized by the emission power \citep{rybicki}
\begin{equation}
\label{syn} 
P_{syn}\simeq\frac{2e^4B^2\gamma^2}{3m^2c^3},
\end{equation}
where $\gamma$ is the particle Lorentz factor, $B\simeq B_{st}\times (r_s/r)^3$ is magnetic field, $B_{st}\simeq 3.2\times 10^{19}\sqrt{P\dot{P}}$ Gauss is the magnetic field close to the surface of the neutron star \citep{shapiro}. Also $P$ ( $\dot{P}$) is the period of rotation (its time derivative), $r_s\simeq 10^6$ cm denotes the star radius,  and $r$ is the radial coordinate.  Throughout the paper we use the $P - \dot{P}$ relation corresponding to the death line (traditionally the radio pulsars are assumed to locate above the so-called death line in $P - \dot{P}$ diagram) $\dot{P}\simeq 10^{-15}\times P^{11/4}$ \citep{ruds}. Therefore, $\dot{P}$ might be even higher, giving a more efficient picture of acceleration. Since the synchrotron induced cooling rate goes up as $\gamma^2$ (Eq. (\ref{syn})), it will eventually balance the rate of energy gain putting a limit on the achieved energy. In the theory of pulsars, it is generally believed that acceleration of particles is very efficient on the length-scales of the light cylinder (LC - a hypothetical area where the linear velocity of rotation equals the speed of light). Applying the synchrotron limiting condition ($dE/dt\simeq P_{syn}$) to electrons in the LC zone, one can obtain an expression for the maximum attainable energy: 
$$\gamma_{syn}^{max}\simeq\left(\frac{dE}{dt}\times \frac{3m^2c^3}{2e^4B^2}\right)^{1/2}\simeq $$
$$\simeq 1.1\times10^{15}\times\left(\frac{f}{10^8\;Hz}\right)^{5/8}\times P^{(2-3\alpha)/16}\times$$
\begin{equation}
\label{gsyn} 
\;\;\;\;\;\;\;\;\;\;\;\;\;\;\;\;\;\;\;\;\;\;\times\;\left(\frac{\dot{P}}{10^{-15} s\;s^{-1}}\right)^{(3\alpha-12)/16}\times\left(\frac{r}{R_{lc}}\right)^{15/4},
\end{equation}
where $\dot{P}$ is normalized by a value typical for normal pulsars ($P\sim 1$ sec), $R_{lc} = cP/2\pi$ represents the LC radius and we have taken into account that the radio spectral luminosity of pulsars, can be approximated as $L_{r,\nu}\simeq 8.5\times 10^3\times\left(B_{st}/P^2\right)^{\alpha}$ erg s$^{-1}$Hz$^{-1}$ with $\alpha\simeq 0.98$ \citep{lumin}. Therefore, it is clear that the resonant mechanism of particle energization is extremely efficient providing relativistic factors of the order of $10^{16}$ on the LC area for the emission frequency $100$ MHz.

 It is also worth noting that any calculation concerning pulsars must take cognizance of the large magnetic fields. Such a calculation will necessarily add effects of order $e\hbar B/(mc E)$ to the energy spectrum, and also cause asymmetry in the wave function. However, given that even for the pulsar fields, the Landau level correction will be, at best, in 10-100 eV range, it is not relevant for the present study.

Curvature Emission could be another mechanism that, in principle, might limit the maximum Lorentz factor. However, the particles, following the curved trajectories, will radiate \citep{curv} if only if co-rotation is maintained. The maximum possible $\gamma$ for which the particles co-rotate is given by  $\gamma_{co-rot}\simeq B_{lc}^2/(8\pi mc^2n_{_{GJ}})$.For normal pulsars ($P\simeq 1$ s, $\dot{P}\simeq 10^{-15} s\; s^{-1}$), it reaches the value $6.5\times 10^6$ on the LC zone. Here $n_{_{GJ}} = {\bf \Omega\cdot B}/(2\pi ec)$ is the so-called Goldreich-Julian number density (or the number density of pulsar's magentospheric particles) \citep{GJ}, and $\Omega = 2\pi/P$ represents the angular velocity of the pulsar. Thus for very high energy particles ($\gamma>> \gamma_{co-rot}$), curvature losses will not be relevant.

Inverse Compton Scattering: The magnetosphere of pulsars is full of thermal photons originating from  the star surface. The temperature of the photon gas depends on the pulsar's age $\tau = P/(2\dot{P})$ \citep{temp}
\begin{equation}
\label{temp2} 
T \simeq \begin{cases}
  5.9\times 10^5\times\left(\frac{10^6\;yrs}{\tau}\right)^{0.1}K  & \tau > 10^{5.2}\; yrs\\
  2.8\times 10^5\times\left(\frac{10^6\;yrs}{\tau}\right)^{0.5}K & \tau > 10^{5.2}\; yrs
\end{cases}
\end{equation} 
that has been obtained by a rough fit to a numerical model presented in \citep{schaab}.

For $P = 1$ sec, $\dot{P} = 10^{-15}$ s s$^{-1}$, $T\simeq 7\times 10^4 K\equiv T_1 $. The accelerated electrons will, inevitably, encounter the thermal photons and lose energy via inverse Compton (IC) scattering. We will consider the IC cooling rates in two different domains, the Compton and the Klein-Nishina regimes. The maximum $\gamma$ acquired by the particle is, as before, attained when the energization rate is balanced by the cooling rate. In the Compton regime, the cooling rate is given by  \citep{rybicki}
\begin{equation}
\label{IC} 
P_{_{IC}}\simeq \sigma_{_{T}}c\gamma^2U,
\end{equation}
(where $\sigma_{_T}$ represents the Thomson cross section, $U\simeq L/(4\pi r^2c)$ is the thermal energy density and $L\simeq 4\pi r_s^2\sigma_{_T}T^4$ - the corresponding luminosity) and would yield  the maximum relativistic factor
$$\gamma_{_{IC}}^{max}\simeq\left(\frac{\bar dE}{dt}\times\frac{4\pi r^2}{\sigma_{_{T}}L}\right)^{1/2}\simeq$$
$$ \simeq 6.9\times 10^6\times\left(\frac{f}{10^8 \; Hz}\right)^{5/8}\times P^{(10-3\alpha)/16}\times$$
\begin{equation}
\label{gic} 
\;\;\;\;\;\;\;\;\;\;\;\;\;\;\times \left(\frac{\dot{P}}{10^{-15 } \; s \; s^{-1}}\right)^{(3\alpha-4)/16}\times\left(\frac{r}{R_{lc}}\right)^{7/4}\times\left(\frac{T_1}{T}\right)^2,
\end{equation}
which comes out to be so high that the condition- $\gamma\epsilon_{ph}/(mc^2)\ll 1$ - necessary for using the formula in the Compton regime is violated ($\epsilon_{ph}\sim kT$  for thermal photons). 

\begin{figure}
  \centering {\includegraphics[width=10cm]{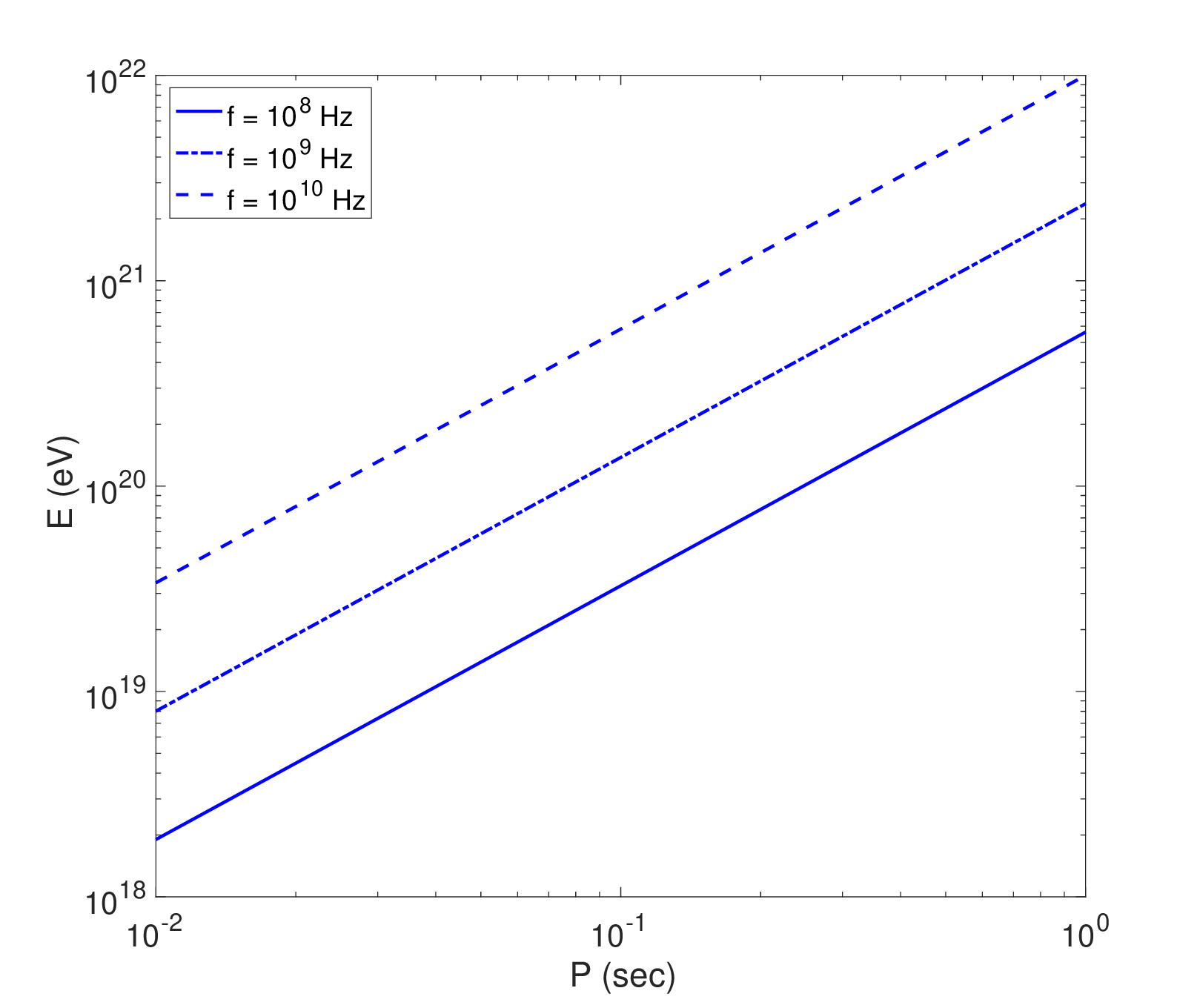}}
  \caption{Here we show the plots of the maximum energy E(P) for different emission frequencies, $f = (1; 10; 100)\times 10^8$ Hz on the LC zone $r = R_{lc}$.}\label{fig1}
\end{figure}
\begin{figure}
  \centering {\includegraphics[width=10cm]{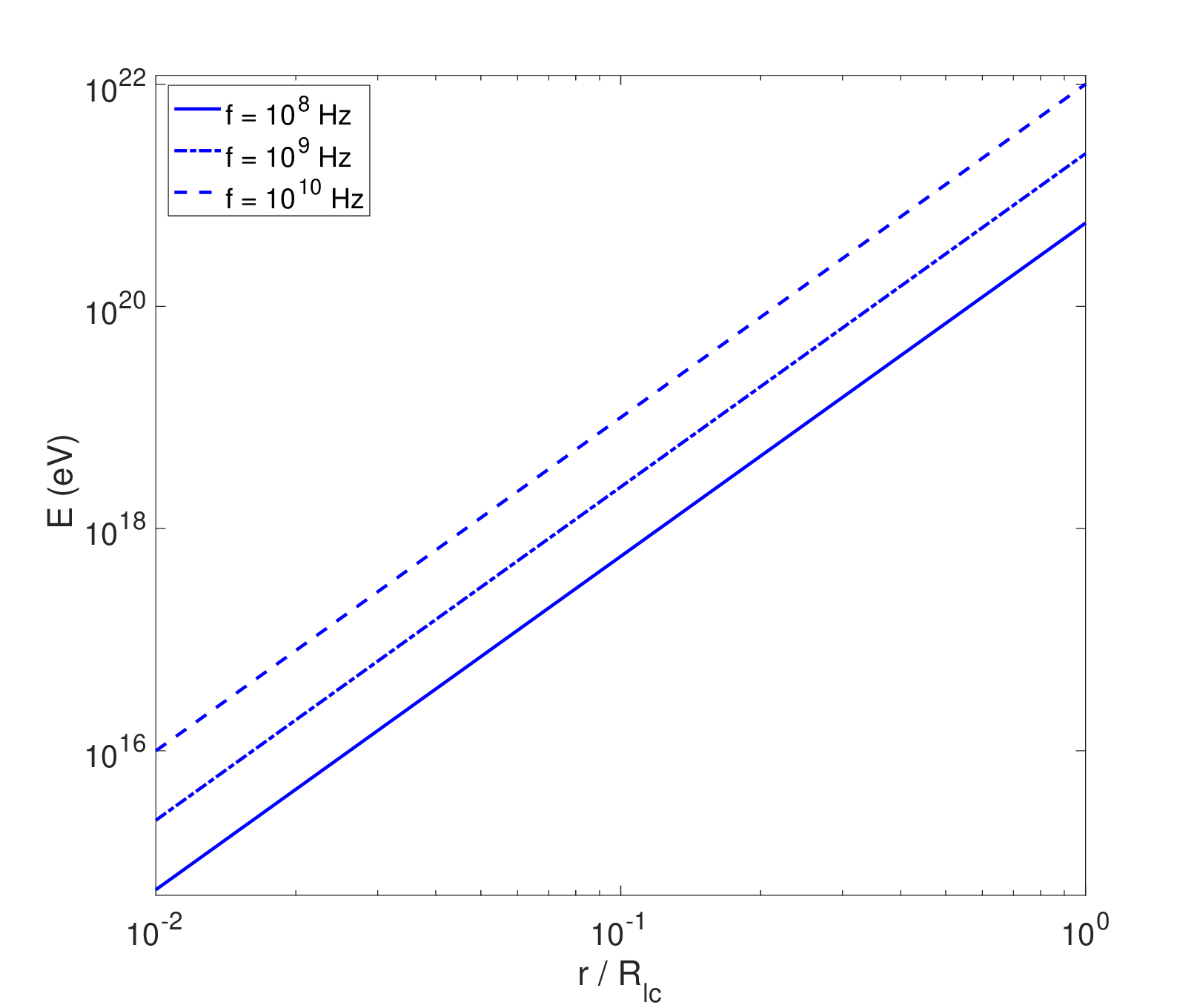}}
  \caption{For $P = 1$ sec we show the maximum energy E(r) versus the radial coordinate for the same set of frequencies, $f = (1; 10; 100)\times 10^9$ Hz.}\label{fig2}
\end{figure}

In the Klein-Nishina regime, characterized by the cooling power \citep{blum} 
\begin{equation}
\label{KN} 
P_{_{KN}}\simeq\frac{\sigma_{_T}\left(mckT\right)^2}{16\hbar^3}\left(\ln\frac{4\gamma kT}{mc^2}-1.981\right)\left(\frac{r_s}{r}\right)^2,
\end{equation} 
a similar prescription (equating $dE/dt\simeq P_{_{KN}}$ ) limits the maximum Lorentz factor to
\begin{equation}
\label{gKN} 
\gamma_{_{KN}}^{max}\simeq 1.5\times10^5\times\frac{T_1}{T}\times\exp\left(1.8\times 10^4\times g\left(r, P,\dot{P}\right)\times\left(\frac{f}{10^8\; Hz}\right)^{5/4}\times\left(\frac{T_1}{T}\right)^2\right),
\end{equation} 

where 
\begin{equation}
\label{g} 
g\left(r,P,\dot{P}\right) = P^{(10-3\alpha)/8}\times\left(\frac{\dot{P}}{10^{-15 } \; s \; s^{-1}}\right)^{(3\alpha-4)/8}\times\left(\frac{r}{R_{lc}}\right)^{3/2}.
\end{equation} 
High values of the factor in the exponent means that the IC process in the KN regime becomes significant only for unrealistically high Lorentz factors. It is worth noting that the so-called Photo-pion reactions, when  photons interacting with protons can produce pions, might become significant, but by taking the corresponding time-scale, $t_{p\gamma}\simeq 10^6\times 10^{19}eV/E$ sec \citep{ahar} into account ($E$ denotes the proton's energy), one can straightforwardly show that the maximum attainable energy is of the order of $10^{28}$ eV, that exceeds by many orders of magnitude the limit corresponding to the synchrotron process. 
Therefore, inverse Compton scattering and the photo-pion processes, like the curvature radiation, do not impose any stringent limits on the maximum allowable energy; the synchrotron cooling process (see Eq. (\ref{gsyn})) is the only relevant one.

In Fig. 1, we plot the behavior of electron energy, $E = \gamma mc^2$ ($\gamma\equiv\gamma_{syn}^{max}$), versus the pulsar period of rotation for three different radio frequencies $f = (1; 10; 100)\times 10^8$ Hz on the LC surface, $r = R_{lc}$. The maximum attainable energy is a continuously increasing function of $P$; resonant energization can catapult electrons to energies of the order of $10^{18-22}$ eV. In the following figure (Fig. 2), for the same set of frequencies and a fixed period of rotation( $P = 1$ sec), we plot $E$ versus the non-dimensional radial coordinate (normalized by $R_{lc})$);  the maximum achievable energy falls in the range $10^{15-22}$ eV.

It is natural to assume that not all of the magnetospheric particles are involved in the energization process. One can estimate a theoretical upper limit on the number density, $n_m$ of such electrons. The total kinetic power of accelerated particles $dK/dt\simeq 2\gamma m n_mc^3\pi r^2\theta^2$ must remain below the slow-down luminosity, $L_{sd} = I\Omega|\dot{\Omega}|=4I\pi^2\dot{P}/P^3$, of the pulsar. In the expression for $L_{sd}$, $I=2Mr_s^2/5$ is the moment of inertia and $M\approx 1.5\times M_{\odot}$ is the pulsar mass ($M_{\odot}\approx 2\times
10^{33}$g is the solar mass). The constraint, $dK/dt<L_{sd}$, puts an absolute upper limit on $n_m$ that varies from $4.4\times 10^{-3}$ cm$^{-3}$ (for $P = 0.001$ sec) to $4.4\times 10^{-4}$ cm$^{-3}$ (for $P = 1$ sec). Realistic number densities will be, surely, much less  because pulsars slow down via the magneto-dipole radiation. Therefore, the flux of ultra-high energy particles will be much less than the theoretical limit, $F\simeq L_{sd}/(4\pi r^2) \sim 4\times 10^{-13}$ erg cm$^{-2}$ s$^{-1}$.

Before summarizing our results, we point a longstanding problem concerning the fate of the low frequency magneto-dipole radiation.  Gunn and Ostriker \cite{gunn1,gunn2} have proposed that the magnetic dipole radiation could accelerate the particles to extremely high energies. This work has been generalized by introducing a refractive index different from vacuum \citep{kegel}, but the details on the working and efficiency of such a mechanism are far from clear \citep{petri}. It may be interesting to explore the interaction of such EM waves with the KG waves.


\section{Summary}
Resonant energization of relativistic electrons via the immense EM fields of a radio pulsar was explored as a mechanism for producing very high energy cosmic rays.  

By considering several "cooling" mechanisms that will eventually compete with resonant energization, we calculated the limiting energies to which the particles could be accelerated. For typical radio pulsars, these limits are set by the synchrotron radiation emission. 

Alternative cooling mechanisms like curvature radiation or Inverse Compton scattering (Thomson or KN regime) on thermal photons,  were found not to impose any significant restrictions on energy gain

The situation could drastically change for the radio-loud pulsars that are gamma-ray sources in the GeV-TeV regime. For such highly energetic photons, IC scattering will become so strong that the particles will not accelerate at all.

When limited, primarily, by synchrotron radiation, the electrons could attain extremely high relativistic factors, $\sim 10^{15}$. This will translate to an energy range $10^{15-22}$ eV for the large range of radio pulsars with periods varying from millisecond to normal values($P = 1$ sec).  

A similar mechanism of acceleration could also pertain for protons, or hadrons; the roles of limiting factors, however, will need to be reassessed. For hadrons, for example, the IC scattering does not impose any constraints on maximum energies \citep{ahar}.

By taking the pulsar's slow-down luminosity into account we have also estimated the upper limit on the number density of electrons that undergo resonant energization;  the fraction of particles that could be so energized, will constitute only an insignificant  fraction of the Goldreich-Julian number density.

\section*{Acknowledgments}\
The work of Z.O. was partially supported by the EU fellowships for Georgian researchers, 2023 (57655523).
Z.O. also would like to thank Torino Astrophysical Observatory and Universit\'a degli Studi di Torino for hospitality during working on this project. SMM acknowledges the support of US DOE grants: DE-FG02-04ER54742 and DE-AC02- 09CH11466.

\section*{Data Availability}

Data are available in the article and can be accessed via a DOI link.

\end{document}